\documentclass[12pt]{article}
\usepackage{epsfig}
\def\slash#1{\setbox0=\hbox{$#1$}#1\hskip-\wd0\hbox to\wd0{\hss\sl/\/\hss}}
\def\bea{\begin{eqnarray}}
\def\eea{\end{eqnarray}}
\def\a{\alpha}
\def\b{\beta}
\def\g{\gamma}
\def\d{\delta}
\def\e{\eta}
\def\ep{\epsilon}
\def\m{\mu}
\def\n{\nu}
\def\l{\lambda}
\def\p{\partial}
\begin{document}
%%%%%%%%%%%%%%%%%%%%%%%%%%%%%%%%%%%%%%%%%%%%%%%%%%%%%%%%%%%%%%%%%%%%%%%%%%%
\thispagestyle{empty}
\begin{flushright}
hep-ph/0307350 \\
July 2003
\end{flushright}
\begin{center}
{\LARGE\bf Updated Collider Bounds on the Parameters \\ 
of Dynamical Torsion} \\
\vspace*{0.4in}
{\large\sl Uma Mahanta$^1$ {\normalsize\rm and} Sreerup Raychaudhuri $^2$} \\
\vspace*{0.3in}
\begin{tabular}{l}
$^1$Harish-Chandra Research Institute, Chhatnag Road, 
Allahabad 211019, India. \\
~E-mail: {\sf mahanta@mri.ernet.in} \\ \\
$^2$Department of Physics, Indian Institute of Technology, Kanpur 208016,
India. \\
~E-mail: {\sf sreerup@iitk.ac.in} \\
\end{tabular}
\end{center}
\vskip 40pt
\begin{center}
{\large\sc Abstract}
\end{center}

\begin{quotation}
\noindent
We constrain the parameters of dynamical torsion, namely, the torsion-fermion 
coupling and the torsion mass, by making a careful analysis of current LEP-2 
data at several energies, as well as CDF and D0 data from Run-I of the
 Tevatron. We find that 
measurements of the forward-backward asymmetry in the $e^+e^- \to \mu^+\mu^-$ 
channel at LEP-2 produce the most restrictive bounds over most of
 the parameter space, 
though other measurements, too, have a significant role to play. 
Our results considerably 
improve the existing constraints on models with dynamical torsion.
\end{quotation}

%%%%%%%%%%%%%%%%%%%%%%%%%%%%%%%%%%%%%%%%%%%%%%%%%%%%%%%%%%%%%%%%%%%%%%%%%%%
\newpage
\section{ Introduction}

The Standard Model (SM) has been extremely successful in explaining 
the results of all current experiments both at low and high energies.  
In spite of this phenomenal success, however, most high energy physicists 
think that the SM is at best an effective field theory valid at low energies
---  one of the main reasons reasons being that the SM does not include 
quantum gravity. At the same time, many theorists believe that the ultimate 
fundamental theory will be provided some day by superstring theory since 
the low-energy effective Lagrangian of closed strings provides a finite
theory of quantum gravity.

The low-energy effective action of closed
 strings  predicts, along with a 
massless graviton, an antisymmetric second rank tensor $B_{\m\n}(x)$
-- the Kalb-Ramond tensor -- which enters the action via its antisymmetrised 
derivative $H_{\m\n\l}= \p_{[\m}B_{\n\l]}$ \cite{GSW}. The third rank
tensor $H_{\m\n\l}$ is referred to as the torsion field strength. Torsion 
is thus a prediction of string theory. It is quite natural, therefore, to 
inquire what observable phenomena at low energies (e.g. collider effects) 
could be produced by the presence of torsion fields in the theory. 
 From the point 
of view of string theory the interaction scale of torsion field
  should be related to and quite close to the string scale $M_s$
 Traditionally one would expect this
string scale to be of the order of the Planck mass $M_{P} \sim 10^{19}$~GeV, 
in which case it is extremely unlikely that laboratory observables would 
have any sensitivity to torsion effects. However Witten has shown \cite{W} 
by explicit calculations in the strong-coupling regime using string duality 
that the string scale can be substantially lower than $M_{P}$. This result 
has been pushed to its extreme by Lykken\cite{L} who speculated that the
string scale could be as low as the electroweak scale -- in which case torsion
effects in laboratory experiments may be as strong as electroweak effects. 
Models of this kind make it particularly attractive to study the effects of 
dynamical torsion at the presently accessible collider energies of a few 
hundred GeV. In fact, it is keeping some such scenario in mind that the 
present study --- a purely phenomenological one --- has been carried out.

Some phenomenological effects at collider energies of heavy, non-propagating 
torsion fields have been considered in the literature\cite{BCHZ}. The
exchange of heavy
torsion fields gives rise to dimension six
four fermion interaction in the low-energy effective action.
Non-propagating torsion therefore has only one unknown parameter
namely the torsion interaction scale. 
Non-observation of deviations from the SM induced by these operators
enables one to obtain a lower bound on the torsion scale $\Lambda_{tor}$.
Dynamical and propagating torsion, on the other hand, has two unknown 
parameters, namely, the torsion mass ($M_S$) and the torsion coupling 
($\eta_S$) to SM fields (fermions). The effects of dynamical torsion on 
precision measurements of forward-backward asymmetries at the CERN LEP-1
experiment, total 
cross-sections at LEP-1.5 and dilepton and dijet production at the Tevatron 
have been considered by Belyaev and Shapiro in Ref.~\cite{BS} and bounds on 
the $\eta_S$--$M_S$ plane have been obtained thereby. However, these 
bounds are somewhat weak as a result of the low statistics available at the 
time their paper was written. Subsequently, both LEP-2 and the Fermilab 
Tevatron have 
acquired enormous amounts of data, all more-or-less confirming the SM 
predictions. It may be expected, therefore, that constraints on new physics 
beyond the SM will be more stringent than
 that before.
 
In this paper, therefore, we update the Belyaev-Shapiro bounds using the
current LEP-2 data and published analyses of CDF and D0 data. In particular, 
we have used the LEP-2 analysis~\cite{LEP} at different energies for the 
forward-backward asymmetry $A^{\mu}_{FB}$ in the $\mu^+\mu^-$ channel and the
data on total cross-section in $e^+e^-\rightarrow e^+e^- (\mu^+\mu^-)$.
From the CDF Collaboration, we have used the dilepton Drell-Yan data 
$p\bar{p}\rightarrow e^+e^- (\mu^+\mu^-)$, and from D0, the dielectron data,
to establish new, updated bounds on the torsion
mass $M_S$ and torsion fermion-coupling $\eta_S$. We find that, as expected, 
our results represent a considerable improvement over the earlier constraints.
We demonstrate that the LEP-2 data on $A^{\mu}_{FB}$ produce the most 
restrictive 
bound over most of the parameter space. However, in the region of low torsion 
mass, where the torsion field can be resonantly produced at the Tevatron, 
the CDF and D0 data yield more stringent constraints.

\section{The Torsion Lagrangian} 

The torsion tensor is defined in terms of the non-symmetric affine connection
$\tilde{\Gamma}^{\a}_{~\b\g}$ by
\bea
T^{\a}_{~\b\g}= \tilde{\Gamma}^{\a}_{~\b\g}- \tilde{\Gamma}^{\a}_{~\g\b} \ .
\eea
This tensor clearly vanishes if the connection is symmetric in $\b\g$,
which is the case with, for example, conventional general relativity.

It is usually found convenient to divide the torsion tensor $T^{\a}_{~\b\g}$ 
into three irreducible components\cite{BS}. These are \\
\hspace*{0.3in} ($a$) a {\it trace}: $T_{\b}=\eta^{\a \g}T_{\a \b \g}$; \\
\hspace*{0.3in} 
($b$) a {\it pseudo-trace}: $S^{\d}= \ep^{\a\b\g\d}T_{\a\b\g}$;  \\
and ~($c$) a third rank tensor $q_{\a\b\g}$: which satisfies the conditions 
    $q^{\a}_{~\b\a} =0$ and $\ep^{\a\b\g\d}q_{\a\b\g}=0$.
\noindent

It is clear that $T_{a}$ behaves as a vector field and $S_{\a}$ as an axial 
vector field. However the simultaneous presence of both $T_{\a}$ and $S_{\a}$,
both of which are coupled to fermions in the low-energy effective Lagrangian,
would lead, in the quantum theory, to serious problems arising from the 
$U(1)_A$ gauge anomaly.
This problem may be circumvented in any one of two different ways. One way 
is to choose the couplings $\eta_{T}(f)$ and $\eta_{S}(f)$ of $T$ and 
$S$ respectively to a fermion pair $f\bar{f}$ so that 
$\sum_f ~C_f ~\eta_{T}^2(f) ~\eta_{S}(f) = 0$, where $C_f$ is a color factor,
in which case the $U(1)_A$ anomaly cancels.
The other way --- which we adopt in this paper --- is to simply assume that 
the torsion tensor possesses a non-trivial pseudo-trace but {\it no} 
non-trivial trace. This is quite sensible from a phenomenological point of 
view. Moreover, such a situation does indeed arise if the torsion tensor is 
antisymmetric in all the three indices.
We thus consider a phenomenological scenario in which the vector field 
$T_\a$ is absent, while the axial vector field $S_\a$ 
interacts with fermions
through an axial-vector coupling. In this case the relevant part of the
Lagrangian can be written 
\bea
{\cal L} = {\cal L}_{SM} - {1\over 4}S_{\m\n}S^{\m\n}
                         +{1\over 2}M^2_S S^{\m}S_{\m}
              + \sum_f \eta_{S}(f) ~\bar{\psi}_f\g_{\m}\g_5\psi_f ~S^{\mu}
\eea
where $S_{\m\n}=\partial_{\m}S_{\n}-\partial_{\n}S_{m}$ and $\psi_f$ 
represents a SM fermion ($f$) field. 
Note that the $U(1)_A$ gauge symmetry associated with the $S$ field
is softly broken by the fermion mass term contained in ${\cal L_{SM}}$ 
and also by the torsion mass $M_S$. Further, in order
that the $U(1)_A$ anomaly associated with $\gamma$--$\gamma$--$S$ and 
$Z$--$Z$--$S$ vertices should cancel, the axial charges
of the SM fermions to the $S$ field may be chosen~\cite{choice} so that 
\bea 
\e_S(u_i) = \e_S(\nu_i) = -\e_S(d_i) = -\e_S(\ell_i) ~~\equiv~ \e_S^i
\label{anomaly}
\eea
where $u_i$ ($d_i$) correspond to quarks of charge $\frac{2}{3}$
(-$\frac{1}{3}$), and $\ell_i$ and $\nu_i$ correspond, respectively,
to charged leptons and neutrinos belonging to the $i^{th}$ generation. 
The above condition still 
allows the value of $\e_S^i$ to change from generation to generation, but
in this paper we 
make the simplifying assumption that $\e_S^i$ is the same for all 
generations i.e. $\e_S^i =\e_S ~\forall i$. It is this generation-independent 
coupling constant $\e_S$ that we constrain using the recent LEP-2 and Tevatron 
data.

\section{Torsion effects on LEP-2 and Tevatron observables}

We have just seen that the pseudo-trace component of the torsion tensor 
effectively behaves as a massive spin-1 field with axial vector couplings
to SM fermions. In many ways,
therefore, its phenomenology is similar to that of an extra 
$Z'$-boson, and its presence will be manifest in the same kind of 
observables which are studied~\cite{Rizzo} in the context of $Z'$ bosons. 
The main observables in question at an $e^+ e^-$ collider like LEP-2 are:

\begin{itemize}
\item The {\it total} cross-section for $e^+ e^- \to f \bar f$, where $f$ is
any SM fermion. This will pick up extra contributions due to diagrams
with $S$-field propagators and the interference of such diagrams with
the SM diagrams. At LEP-2, data are available~\cite{LEP} for the $\mu^+\mu^-$,
$\tau^+ \tau^-$, Bhabha ($e^+e^-$) and dijet channels. While Bhabha
scattering is an obvious first choice because of the large cross-section,
data for the $\mu^+\mu^-$ channel are also clean and have small error-bars 
and can therefore be used very effectively. Data for the $\tau^+ \tau^-$ 
and dijet channels are less useful because of the larger error-bars.

\item The {\it differential} cross-section for $e^+ e^- \to f \bar f$.
It is well known that the presence of both vector and axial vector couplings
in the effective Lagrangian would lead to parity violating signatures
in collider experiments. The most convenient variable where it shows up
is the precision measurement of forward-backward asymmetry.
 A source of $A_{FB}^f$
already exists in the SM because of the $\g_{\m}\g_5$ component
of $Z$ and $W$-boson interactions with fermions. The presence of torsion
with axial-vector type interactions will lead to additional 
parity-violating effects and change the forward-backward asymmetry from its
SM value. It is not advisable to use
the forward-backward asymmetry data for Bhabha scattering $e^+e^- \to e^+e^-$, 
because the large $t$-channel photon contribution creates a very large
forward-backward asymmetry, which is highly sensitive to the angular cuts 
used in the analysis, and tends to overwhelm genuinely parity-violating effects.
\end{itemize}
An important possibility at LEP-2 is the fact that $s$-channel exchanges
of the $S$-boson could lead to resonances in the total cross-section. For
torsion masses accessible at LEP-2 energies, this would lead to enormous
enhancements in the cross-section and the simple fact of their non-observance 
can be used to put stringent bounds on the torsion coupling $\eta_S$ for
the kinematically accessible region in the parameter space. However,
virtual torsion fields can also contribute to LEP-2 observables. As
the measured values of almost all observables at LEP-2 are known to be in 
excellent agreement with their SM predictions, it follows that one can 
establish bounds on torsion parameters over their entire range, subject only 
to the limitations imposed by experimental errors.

The situation at the Tevatron is generally similar, but with some important 
differences. We can look for torsion contributions to 
$q\bar q \to f \bar f$, where the 
observable final state would have dileptons or dijets. However, it is not
feasible to study dijets because the enormous QCD background 
would tend to swamp any torsion effects. We, therefore,
limit our study to Drell-Yan dileptons only, and, in particular, to 
($a$)~the CDF data~\cite{CDF} on $p \bar p \to e^+ e^- (\mu^+ \mu^-)$ and 
($b$)~the D0 data~\cite{D0} on 
$p \bar p \to e^+ e^-$. In every case, the $S$-boson contribution arises 
through
an $s$-channel propagator, as in the case of $e^+e^- \to q \bar q$.
Since the energy of the initial-state partons is variable, some of the
events are always sure to hit an $S$-resonance, should it be kinematically 
accessible. This leads to large torsion contributions, and hence --- since
the observed data match the SM predictions very well --- to strong constraints.
As the torsion mass $M_S$ increases, the Bjorken variables $x_1, x_2$
required for resonant production go up and the consequent steep fall in parton
luminosities kills the cross-section, thereby weakening the
constraints. In each case, we have considered the bin-wise invariant mass 
distribution only. The CDF Collaboration has also presented~\cite{CDF} 
forward-backward
asymmetry measurements in $p \bar p \to e^+ e^-$, but these data have large
errors which are traceable to the low charge-identification efficiencies for
$e^\pm$. Hence, we have elected not to use this data for our analysis.
The D0 Collaboration has presented the invariant mass distribution 
for dielectrons, but were unable to make forward-backward asymmetry 
measurements since their detector lacks a central magnetic field~\cite{D0}. 
Thus, a similar analysis is required for all three sets of experimental 
data from the Tevatron.

\section{Torsion constraints from LEP-2 data} 

In this paper we have focused on the following measurements:
\begin{enumerate}
\item The total cross-section $\sigma(e^+e^- \to \mu^+\mu^-)$, which is 
compared with the SM value $\sigma_{SM}$ by studying the ratio 
$\sigma/\sigma_{SM}$.
\item The forward-backward asymmetry in the muon channel $A_{FB}^\mu$,
which is compared with the SM value $A_{FB}^{\mu(SM)}$ by studying the
difference $A_{FB}^\mu - A_{FB}^{\mu(SM)}$.
\item The total cross-section $\sigma(e^+e^- \to e^+e^-)$ for Bhabha
scattering.
\end{enumerate}

\bigskip
%---------------------------------------------------
\begin{figure}[htb]
\begin{center}
\vspace*{3.5in}
      \relax\noindent\hskip -4.8in\relax{\includegraphics{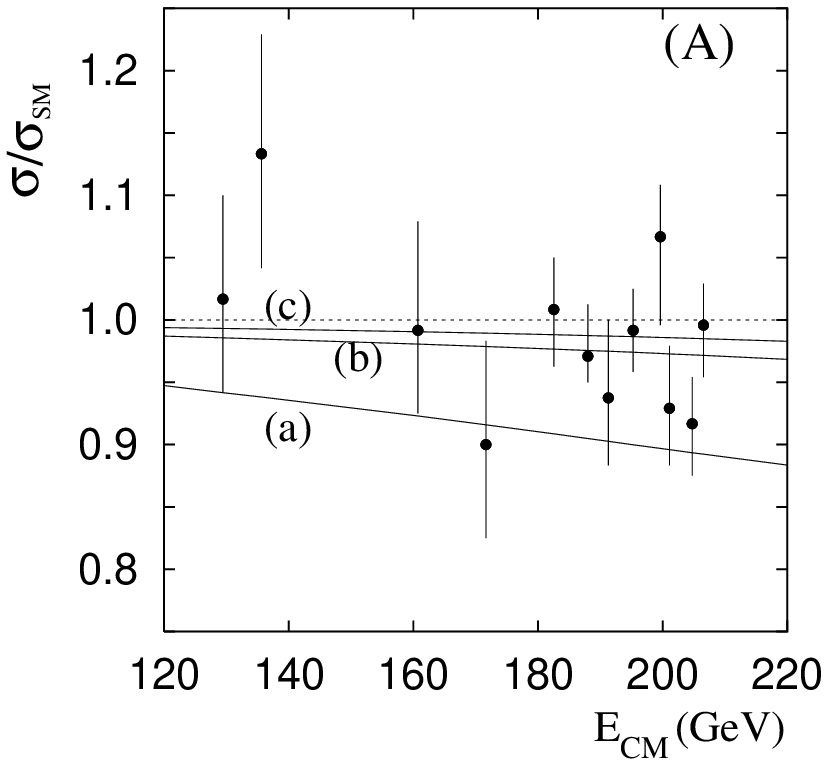}}
      \relax\noindent\hskip -2.8in\relax{\includegraphics{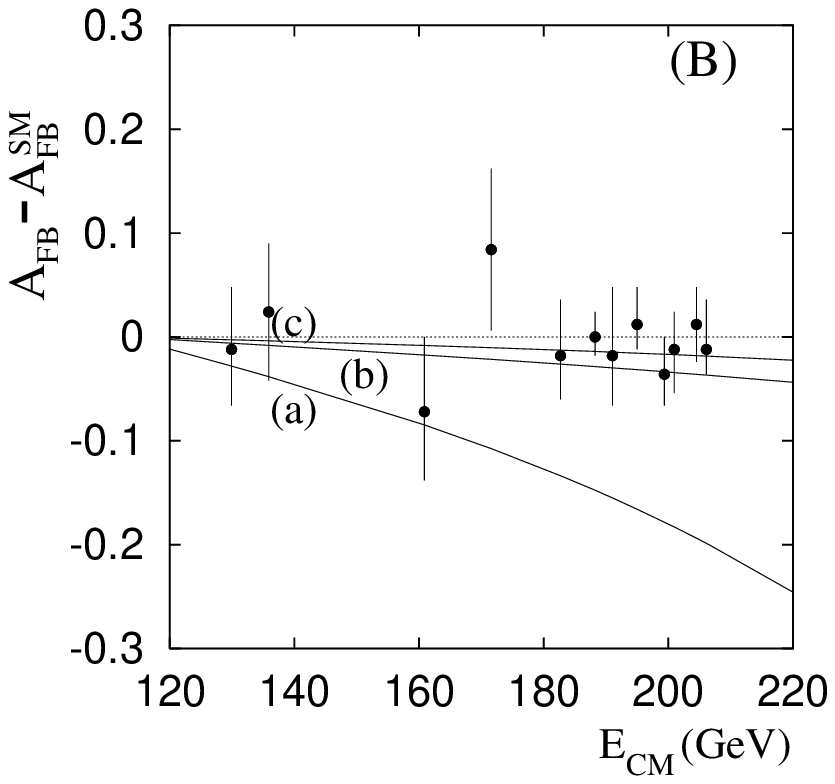}}
\end{center}
\vspace*{-0.8in}
\caption{\footnotesize\sl 
Illustrating the torsion contributions to the process
$e^+e^- \to \mu^+ \mu^-$ at LEP-2. We plot ($A$) the ratio 
$\sigma/\sigma_{SM}$ and ($B$) the difference 
$A_{FB}^\mu - A_{FB}^{\mu(SM)}$. The dotted line represents the SM 
prediction; solid lines represent the torsion effects for
($a$)~$M_S = 500$~GeV; $\e_S = 0.3$, 
($b$)~$M_S = 1$~TeV; $\e_S = 0.3$, and 
($c$)~$M_S = 500$~GeV; $\e_S = 0.1$.
The points with error-bars represent data from LEP-2.
}
\end{figure}
%---------------------------------------------------
\vskip 5pt

The exact form in which these variables are taken is determined by the 
experimental data 
presented~\cite{LEP} by the LEP-2 Collaborations at DPF-2002. The
use of $\sigma/\sigma_{SM}$ is well-known to cancel out the principal 
effects due to initial state radiation. The use of 
$A_{FB}^\mu - A_{FB}^{\mu(SM)}$ isolates the new physics contribution
to forward-backward asymmetry.  In Figure 1 we present a
comparison of torsion effects with the SM prediction and the LEP-2
data for ($A$) $\sigma/\sigma_{SM}$ and ($B$) $A_{FB}^\mu - A_{FB}^{\mu(SM)}$ 
respectively. The dotted line represents the SM predictions (1 and 0
respectively), while solid lines represent the results of including
torsion effects for \\
\hspace*{0.3in} ($a$) $M_S = 500$~GeV; $\e_S = 0.3$, \\
\hspace*{0.3in} ($b$) $M_S = 1$~TeV; $\e_S = 0.3$,  \\
and ~($c$) $M_S = 500$~GeV; $\e_S = 0.1$. \\
The points with error-bars represent data from LEP-2 at different energies.
It is immediately clear that the data are  
usually within $1\sigma$ of the SM value. It is also clear that 
substantial deviations from the SM prediction
occur only if the torsion mass is relatively light and the torsion-fermion
coupling is close to electroweak strength. Even then, the total 
cross-section shows a relatively small deviation; 
for the forward-backward asymmetry,
however, for energies above around 180 GeV, the deviation can be considerable.
It follows that we can derive a strong bound using the $A_{FB}^\mu - 
A_{FB}^{\mu(SM)}$ data. In order to do this, the general procedure
adopted has been to calculate a $\chi^2$ for the data on a variable $Q$
using the formula
\bea
\chi^2(\e_S,M_S) = \sum_i \frac{[Q_i^{th}(\e_S,M_S) - Q_i^{c.v.}]^2}
{(\delta  Q_i^{exp})^2}
\label{chisq}
\eea
where $i$ runs over the different values of $E_{CM}$ at LEP-2, 
$Q_i^{th}(\e_S,M_S)$ is the theoretically-computed value, 
$ Q_i^{c.v.}$ is the experimental central value and $\delta  Q_i^{exp}$
represents the experimental error. As the errors presented by
the LEP-2 collaborations are asymmetric, we choose the error-bar on the 
same side of
the central value as the theoretical curve. Constraints may now be
obtained by varying $\e_S,M_S$ and demanding that the resultant $\chi^2$
should not exceed the value permissible for random Gaussian variables.
This has, in fact been done in Figure~2, where
we constrain the $\eta_S$--$M_S$ plane using these two measurements. It is
obvious from Figure~2 that the forward-backward asymmetry measurement is the 
more useful one in constraining torsion parameters. We attribute this to
the fact that the $A_{FB}^\mu$ parameter is not affected by the large 
QED contribution.

%---------------------------------------------------
\begin{figure}[htb]
\begin{center}
\vspace*{3.95in}
      \relax\noindent\hskip -4.8in\relax{\includegraphics{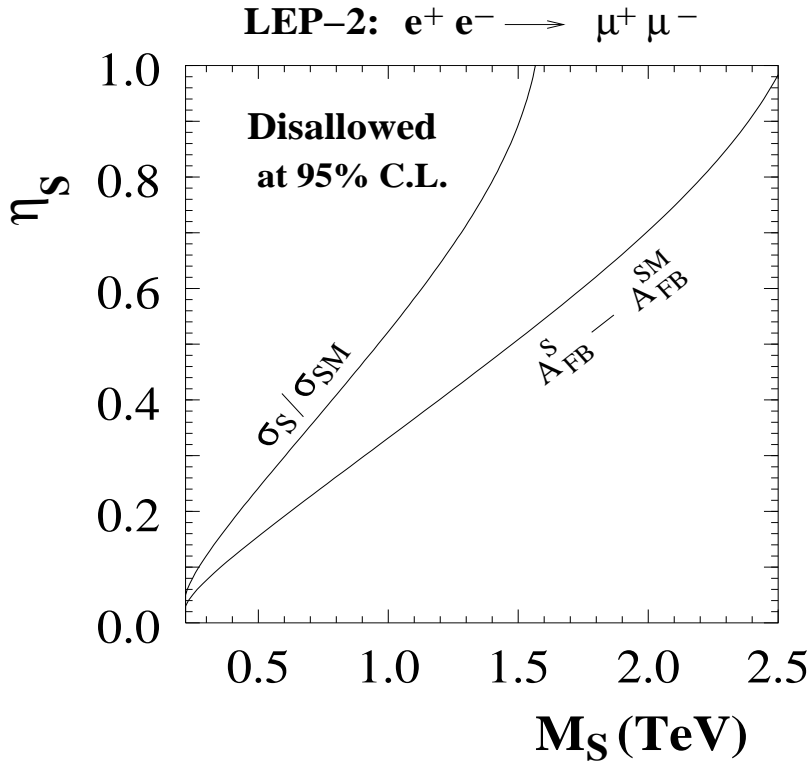}}
\end{center}
\vspace*{-0.4in}
\caption{\footnotesize\sl
Illustrating LEP-2 constraints on the parameter space obtained from 
data on $e^+e^- \to \mu^+ \mu^-$.
}
\end{figure}
%---------------------------------------------------
\vskip 5pt

We now turn to Bhabha scattering, which is a somewhat more complicated
calculation, given the fact that there are $t$ as well as $s$-channel 
contributions from photon, $Z$-boson and $S$-boson propagators, with 
interference
terms between all the contributions~\cite{BP}. Figure 3($A$) shows the excess
contributions compared with the SM (dotted line) as solid lines for \\
\hspace*{0.35in} ($a$) $M_S = 300$~GeV; $\e_S = 0.3$, \\
\hspace*{0.35in} ($b$) $M_S = 500$~GeV; $\e_S = 0.3$, \\
and ~($c$) $M_S = 800$~GeV; $\e_S = 0.3$, \\
where, the points with error-bars represent, as before, data from LEP-2 
at different energies~\cite{LEP}. Clearly, one obtains substantial deviations
only for lighter values of $M_S$. Making a $\chi^2$ analysis as before, we 
obtain constraints on the parameter space illustrated in Figure 3($B$).
Because of the small error-bars in Bhabha scattering data, the constraints
arising from this rival those from the forward-backward asymmetry, at
least up to
about $M_S \approx 1.7$~TeV, after which the Bhabha scattering process
shows little or no contribution from $S$-bosons.

\bigskip
%---------------------------------------------------
\begin{figure}[htb]
\begin{center}
\vspace*{3.6in}
      \relax\noindent\hskip -4.8in\relax{\includegraphics{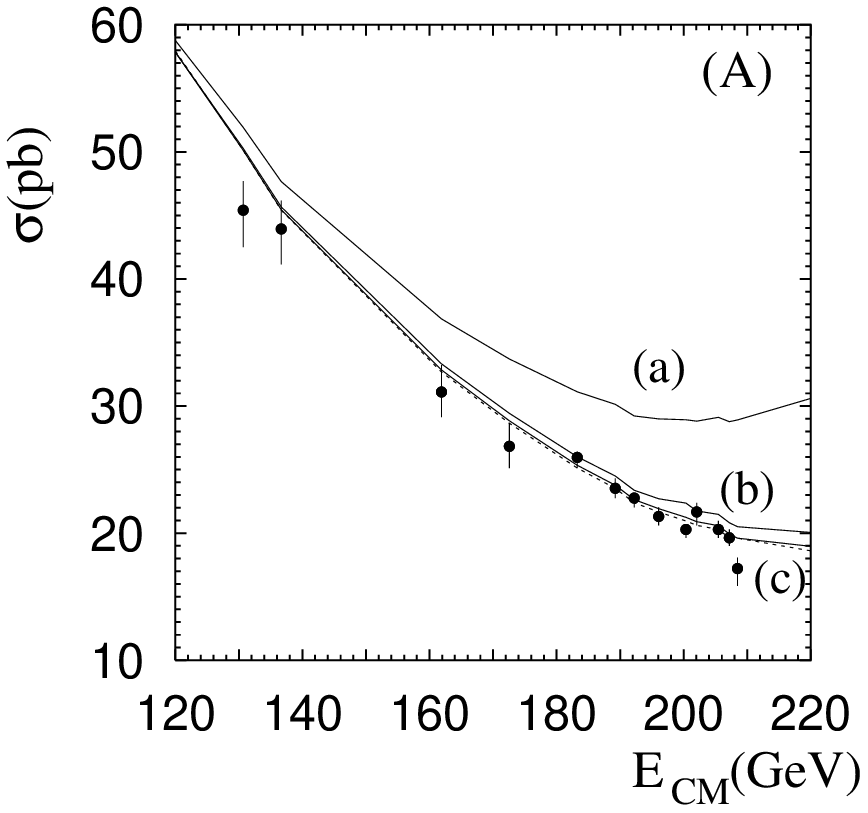}}
      \relax\noindent\hskip -2.8in\relax{\includegraphics{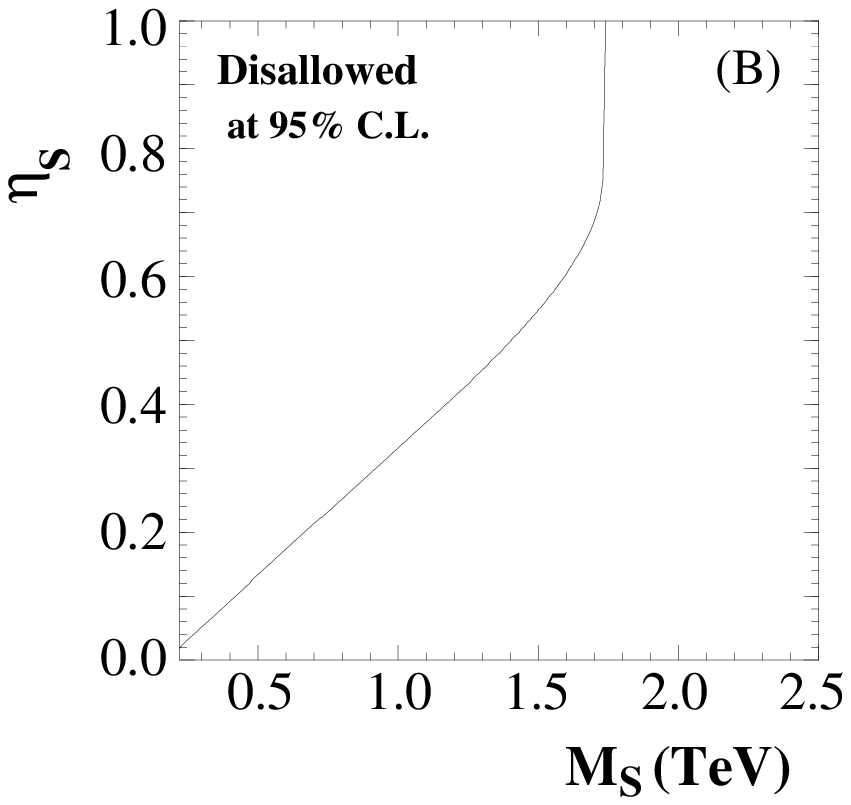}}
\end{center}
\vspace*{-0.8in}
\caption{\footnotesize\sl
Illustrating the torsion contributions to the Bhabha scattering process
$e^+e^- \to e^+ e^-$ at LEP-2. ($A$) shows the cross-section:
the dotted line represents the SM
prediction, solid lines represent the torsion effects for
($a$) $M_S = 300$~GeV; $\e_S = 0.3$, 
($b$) $M_S = 500$~GeV; $\e_S = 0.3$, and 
($c$) $M_S = 800$~GeV; $\e_S = 0.3$, while
the points with error-bars represent data from LEP-2. The corresponding
95\% C.L. constraints on the parameter space are shown in ($B$).
}
\end{figure}
%---------------------------------------------------
\vskip 5pt

If we compare these bounds with those obtained earlier~\cite{BS} using LEP-1.5
data, we notice an enormous improvement, which is primarily due to the 
fact that the error-bars decrease at higher LEP 
energies: a result of the higher luminosity accumulated at
these energies. The smaller error-bars are most significant in Figure 3($A$).
In any case, it may be argued that the higher the energy, the closer we are 
to a heavy $S$-resonance and the larger must be the corresponding torsion 
contribution. Hence we should expect even more stringent bounds at a 
500~GeV or 1~TeV linear
$e^+e^-$ collider or, perhaps, at a high energy muon collider where 
center-of-mass energies as high as 3--4~TeV or even as high as 10~TeV 
have been contemplated~\cite{MuonC}.

\section{Torsion constraints from Tevatron data} 

At the Tevatron collider, running (Run-I) at a center-of-mass energy of 
1.8~TeV, we would expect to see torsion resonances in $p \bar p \to e^+ e^- 
(\mu^+ \mu^-)$, as 
explained earlier. The best observables where this would show up would 
be in the bin-wise invariant mass distribution given, for example, by
both CDF and D0 Collaborations, where a distinct peak would appear at 
the mass of the resonance. We have calculated the parton-level cross-sections
keeping in mind different signs for the $S$-boson couplings to $u$ and $d$
quarks, according to the convention of Eqn.~\ref{anomaly}.
The CDF and D0 data are reproduced in Table 1. 
For our numerical analysis of the CDF $e^+ e^-$ data, the parton-level 
cross-section with torsion 
effects included is incorporated into a Monte Carlo event generator, 
running subject to the following kinematic cuts:
\begin{itemize}
\item The final state electron (positron) should have pseudorapidity
within $|\e_e|<4.2$ to be detected in the electromagnetic calorimeter.
The pseudorapidity coverages in the CDF detector are: $|\e_e|<1.1$ for
the central calorimeter, $1.1<|\e_e|<2.4$ for the end plug, and 
$2.2<|\e_e|<4.2$ for the forward calorimeter.
\item At least one final state electron (positron) should have pseudorapidity
$|\e_e|<2.4$ so that it does not lie in the forward calorimeter.
\item The minimum transverse momentum $p_{Te}$ of the electron (positron)
should be 22, 20 or 15~GeV, depending on whether it passes into the
central calorimeter, end cap or forward calorimeter.
\item The electron (positron) tracks should not be back-to-back. This 
reduces the cosmic ray background. To implement this we demand that
the dielectron opening angle should satisfy $\theta_{ee} < 175^0$.
\end{itemize}

These kinematic cuts are identical with the ones used by the CDF
Collaboration in their analysis~\cite{CDF}, 
so far as they can be translated to a 
parton-level analysis. Our numerical results compare well with the 
experimental numbers given in the first part of Table~1. 

For a dimuon final state, the kinematic cuts are much simpler. We demand
\begin{itemize}
\item The final state muons should have pseudorapidity $|e_\m| < 1.1$, which
is the coverage of the muon chambers.
\item The final state muons should have a minimum transverse momentum
$p_{T\mu} > 2.8$~GeV, which is required for triggering. 
\item The dimuon opening angle should satisfy $\theta_{\mu\mu} < 175^0$,
to eliminate cosmic ray backgrounds.
\end{itemize}

%---------------------------------------------------
\footnotesize
\begin{center}
\begin{tabular}{|cccccccccc|}
\hline
CDF & $e^+e^-$ & & & & & & & & \\ \hline
$M_{e^+e^-}$ &
105--120 & 120--140 & 140--200 & 200--300 & 300--400 & 400--600 & 600--999 & 
& \\
$\sigma$ (pb)& ~3.934 & ~1.309 & ~1.249 & ~0.260 & ~0.081 & ~0.028 & ~0.000 & 
& \\
$\delta \sigma$ (pb) & $\pm 0.355$ & $\pm 0.203$ & $\pm 0.192$ & $\pm 0.085$ & 
$\pm 0.047$ & $\pm 0.028$ & $\pm 0.036$ & & \\
\hline  \hline
CDF & $\mu^+\mu^-$ & & & & & & & &\\ \hline
$M_{\mu^+\mu^-}$ &
110--120 & 120--150 & 150--200 & 200--300 & 300--400 & 400--600 & 600--999 & 
& \\
$\sigma$ (pb)& ~2.319 & ~2.418 & ~0.699 & ~0.391 & ~0.056 & ~0.000 & ~0.000 & 
& \\
$\delta \sigma$ (pb) & $\pm 0.516$ & $\pm 0.478$ & $\pm 0.233$ & $\pm 0.158$ & 
$\pm 0.056$ & $\pm 0.040$ & $\pm 0.044$ & & \\ 
\hline \hline 
D0 & $e^+e^-$ & & & & & & & &\\ \hline
$M_{e^+e^-}$ &
120--160 & 160--200 & 200--240 & 240--290 & 290--340 & 340--400 & 400--500 & 
500--600 & 600--1000 \\ 
$\sigma$ (pb)& ~1.930 & ~0.490 & ~0.280 & ~0.066 & ~0.033 & ~0.057 & ~0.039 &
~0.037 & ~0.035 \\
$\delta \sigma$ (pb) & $^{+0.430}_{-0.440}$ & $^{+0.160}_{-0.180}$ &
$^{+0.090}_{-0.100}$ & $^{+0.052}_{-0.058}$ & $^{+0.032}_{-0.030}$ &
$^{+0.042}_{-0.047}$ & $^{+0.024}_{-0.039}$ & $^{+0.023}_{-0.037}$ &
$^{+0.023}_{-0.035}$ \\
\hline \hline
\end{tabular}
\end{center}
\normalsize
\noindent
Table 1: {\footnotesize\sl Drell-Yan data from the Tevatron showing the 
observed cross-section with error-bars deposited in different bins of 
dilepton invariant mass. 
The CDF and D0 data are taken from Refs.~\cite{CDF} and Ref.~\cite{D0}
respectively. All invariant masses are measured in GeV.}
%---------------------------------------------------
\vskip 10pt

With these cuts our numbers agree well with the second part of Table~1.
For parton fluxes we have followed the CDF Collaboration in using the 
MRST99 structure functions~\cite{MRST}, 
and we have also incorporated next-to-leading-order QCD corrections in the
same way as the CDF Collaboration have, namely, by weighting the events 
with a $K$-factor~\cite{Altarelli}
\bea
K(\hat{s}) = 1 + \frac{4}{3} \left( 1 + \frac{4}{3}\pi^2) \right)
\frac{\alpha_s(\hat{s})}{2\pi} \ .
\eea

The D0 Collaboration has presented~\cite{D0} 
a similar set of $e^+e^-$ data, as shown 
in the last part of Table~1. To compare our predictions with their data, 
we use the following set of kinematic cuts.
\begin{itemize}
\item The final state electron (positron) must lie within the angular
coverage of the electromagnetic calorimeter, i.e. it must have either
$|\e_e|<1.1$ to be in the central calorimeter (CC) or $1.5<|\e_e|<2.5$ 
to be in the forward calorimeters (EC).
\item The final state electron (positron) must have a minimum transverse 
energy $E_{Te}>20$~GeV.
\item The dielectron opening angle should satisfy $\theta_{ee} < 175^0$
in order to eliminate cosmic ray backgrounds.
\end{itemize}
QCD corrections are included by the inclusion of a $K$-factor.
Following the D0 Collaboration, we make a numerical calculation of
the $K$-factor by comparing our LO results for the SM cross-section
with the NNLO calculations of Ref.~\cite{Neerven}. We have used
MRS(A$'$) structure functions~\cite{MRSA}, again following the D0 Collaboration.
With these included, our cross-sections match well with the D0 data
given in the last part of Table~1. 

%---------------------------------------------------
\begin{figure}[htb]
\begin{center}
\vspace*{4.0in}
      \relax\noindent\hskip -4.8in\relax{\includegraphics{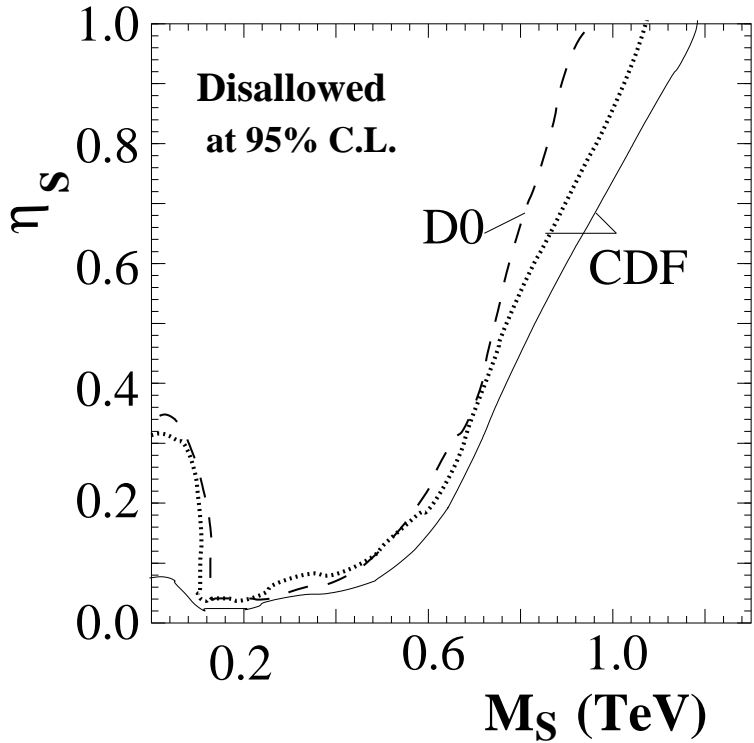}}
\end{center}
\vspace*{-0.3in}
\caption{\footnotesize\sl
95\% C.L. constraints on the parameter space obtained from Drell-Yan data
measured in Run-I of the Tevatron. The solid (dotted) curve corresponds to
dielectron (dimuon) final states at the CDF detector, while the dashed
curve corresponds to dielectron final states at the D0 detector.
}
\end{figure}
%---------------------------------------------------
\vskip 5pt

Using the above tools, we are now in a position to generate the
invariant mass distribution for the torsion-inclusive theory and
compare it with the CDF and D0 data.
Since we are handicapped by not knowing the actual value of
the torsion mass, it is sensible to make a $\chi^2$ analysis of the
invariant mass distribution using the same formula as in Eqn.~(\ref{chisq}),
except that the sum now runs over invariant mass bins rather than $E_{CM}$
values. Once again, requiring that the calculated $\chi^2$ for a given
pair of ($\eta_S$, $M_S$) should not exceed the maximum value permissible
for random Gaussian fluctuations, we obtain constraints on the
$\eta_S$--$M_S$ plane. Clearly, the presence of a resonance would
send the $\chi^2$ shooting up; on the other hand, the result will
be somewhat diluted by the inclusion of the other mass bins, where the
deviation will be minimal. Accepting these features of the statistical
method, we present our
results in Figure 4. It is clear that torsion masses up to about 600~GeV
are very strongly constrained by the Drell-Yan data; beyond this, as has
been explained, the rapid fall in parton fluxes weakens the bound
considerably, so that, in fact, it is no longer competitive with LEP-2 bounds.

It is also clear that the CDF dielectron data provide marginally stronger 
constraints on torsion parameters than the CDF dimuon and D0 dielectron data.
The first is not unexpected, as the CDF dielectron data have smaller 
error-bars than the dimuon data. On the other hand, the D0 dielectron data, 
though having comparable error-bars, yield a slightly poorer constraint.
We attribute this to the marginally larger number of bins (9 instead of 7), 
which creates a greater dilution effect in our $\chi^2$ analysis, as explained 
above. The difference between the solid and dashed curves in Figure~4 is, 
therefore, primarily an artifact of our somewhat naive statistical method. 
As all the three data sets produce results in the same ballpark, however, 
we do not consider it worthwhile, at the present stage, to try a more 
sophisticated statistical method~\cite{Bayesian} for the D0 constraints. 
Moreover, we shall show in the next section that in the region where the
three curves differ, a stronger constraint is provided anyway by the LEP-2 
data analyzed in the previous section. 

\section{Combined bounds} 

In Figure 5, we have shown the constraints
on the $\eta_S$--$M_S$ plane by combining the most significant of the 
constraints obtained above. These are the following: 

\noindent
\hspace*{0.3in} ($a$) constraint from $\sigma/\sigma_{SM}$ for 
$e^+e^- \to \mu^+\mu^-$ at LEP-2; \\
\hspace*{0.3in} ($b$) constraint from $A_{FB}^\mu - A_{FB}^{\mu(SM)}$ at LEP-2; \\
\hspace*{0.3in} ($c$) constraint from Bhabha scattering at LEP-2; \\
\hspace*{0.3in} ($d$) constraint from $p\bar p \to e^+e^-$ at the 
Tevatron (CDF). 

Obviously, constraint ($b$) is the most effective of the LEP-2 constraints,
except for a tiny sliver of parameter space between 600-850~GeV, where
Bhabha scattering is marginally more restrictive. Below 600~GeV, however,
the strongest constraints come from the CDF dielectron data ($d$). We
have not included the CDF dimuon data or the D0 dielectron data because
the CDF dielectron data turn out to be more effective in constraining the 
torsion parameters.

Taken all together,
we rule out a considerable area in the parameter space (roughly double
the area ruled out in the earlier analyses). Among other things, it is
clear that if the torsion coupling is to be of electroweak strength
($\eta_S \sim 0.3 - 0.6$), then the torsion mass will be pushed up
well above a TeV. If, on the other hand, the torsion-fermion coupling
is weak ($\eta_S \sim 0.01$), there is practically no laboratory bound
and it would still be possible to have very light excitations of the 
torsion field.    

\bigskip
%---------------------------------------------------
\begin{figure}[htb]
\begin{center}
\vspace*{3.0in}
      \relax\noindent\hskip -4.8in\relax{\includegraphics{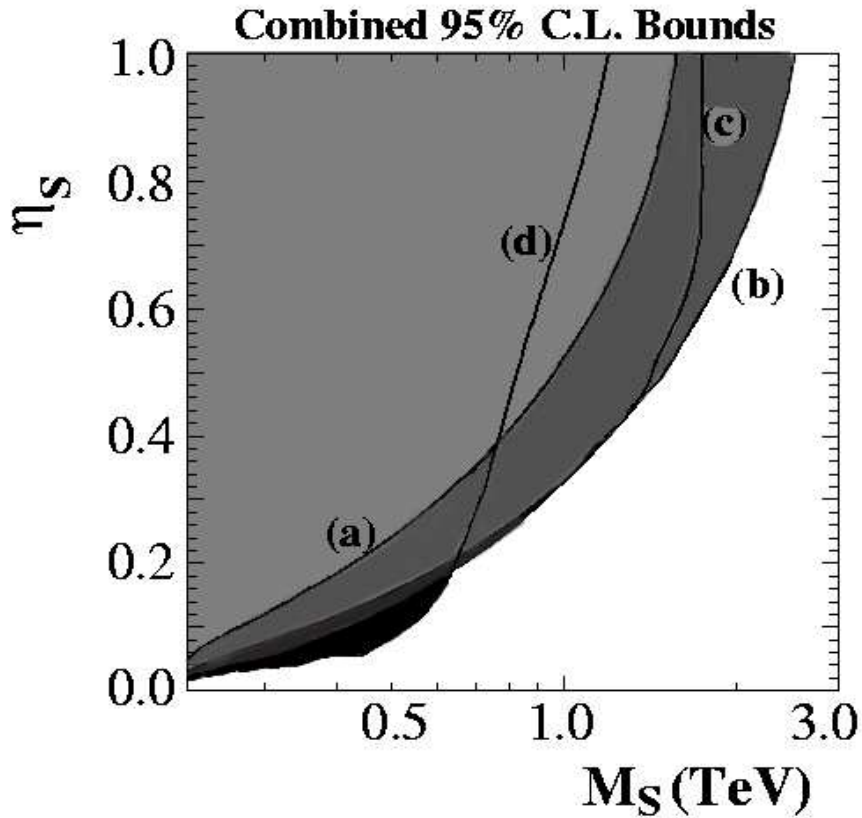}}
\end{center}
\vspace*{0.9in}
\caption{\footnotesize\sl
Combined constraints on the parameter space obtained from LEP-2 and the
Tevatron. These 95\% C.L. curves correspond to
($a$) $\sigma/\sigma_{SM}$ and ($b$) $A_{FB}^\mu - A_{FB}^{\mu(SM)}$ 
for $e^+e^- \to \mu^+\mu^-$ at LEP-2; 
($c$) Bhabha scattering at LEP-2; 
($d$) the CDF measurement of $p\bar p \to e^+e^-$ at the Tevatron. 
}
\end{figure}
%---------------------------------------------------
\vskip 10pt

At Run-II of the Tevatron, which is currently in progress, we expect some
improvement in the bounds because of two circumstances. In the first
place, the center-of-mass energy is marginally higher (2~TeV) and hence
the kinematic reach for resonant $S$-bosons also increases marginally.
Within this region, the accumulation of high luminosity could strengthen
the constraint on the coupling $\e_S$. This would also better the present
bound on $\e_S$ in the non-resonant region, but not, perhaps, very 
significantly. It is unlikely, therefore, that Run-II constraints
would be competitive with LEP-2 results when the torsion mass is of the order
of a TeV or more.

\section{Discussion and Conclusions}

Before we conclude, we need to take note of some important points.
We have assumed that the coupling $\e_S$ is universal, i.e. the same
for every generation. If this assumption is relaxed, the constraints
from Bhabha scattering and from Dell-Yan dielectrons become constraints
on the first generation coupling $\e_{S1}$, while those from 
$e^+e^- \to \mu^+ \mu^-$ and from Drell-Yan dimuons become constraints
on the combination $\sqrt{\e_{S1}\e_{S2}}$. Combining the two could
could yield constraints on $\e_{S2}$ alone. In a similar way, we could
derive (weaker) constraints on $\e_{S3}$ by using the LEP-2 data on
$e^+e^- \to \tau^+ \tau^-$. However, such an exercise would become
worthwhile only if there is some {\it empirical} reason to suspect that
the torsion couplings could change from generation to generation.

In our analysis of collider bounds, we have focused on the effects
of light dynamical torsion with weak coupling $\eta_S \leq 0.1 - 1.0$ to 
fermions. Our results, as shown in Figure 5, demonstrate that with 
increasing $\eta_S$,
the lower bound on $M_S$ is driven to higher values. It follows that for
strong torsion-fermion couplings (e.g. $\eta_S \simeq \sqrt{4\pi}$) the
results obtained with high mass dynamical torsion would more-or-less agree
with those of non-propagating torsion, which are already available in the
literature~\cite{BCHZ}.

In this paper we have considered the collider effects of the 
pseudo-trace component $S_{\mu}$ of the torsion tensor only and
assumed that the trace field $T_{\mu}$ vanishes. However
it is also possible to consider a scenario in which the torsion tensor
is symmetric in any two indices. In this scenario $S_{\mu}$ vanishes but 
$T_{\mu}$ has a non-trivial value. Since $S_{\mu}$ is an axial vector
but $T_{\mu}$ is a vector, the forward-backward asymmetry that they produce
are expected to differ significantly. In the case of $S_{\mu}$ the additional
$A_{FB}$ due to torsion arises from the photon-$S_{\mu}$
interference and the interference between $S_{\mu}$ and the vector coupling
of $Z_{\mu}$. However since the vector coupling of $Z_{\mu}$
to charged leptons is very small the later contribution is expected to be
small except near the $Z$-pole or $S$-pole. On the other hand in the case of
$T_{\mu}$ the additional contribution to $A_{FB}$ arises
from the interference between $T_{\mu}$ and the axial vector coupling
of $Z_{\mu}$. Since the axial coupling of $Z_{\mu}$ to charged leptons
is quite large, the later contribution is expected to be appreciable even
away from $Z$ and $S$-boson pole. The bounds in the $\eta_S -M_S$ plane 
 arising from current collider data are therefore expected to differ
from the case of $S_{\mu}$.

To sum up, then, we have updated the collider bounds on dynamical torsion
in a scenario where only the pseudo-trace field $S_\mu$ couples to fermions,
with a universal coupling.
Using current data from LEP-2 and the Tevatron, we find that torsion fields
up to about 600 GeV are excluded unless the torsion-fermion coupling drops
below 0.1. This is a much stronger constraint than those obtained from 
considering either the anomalous magnetic moment of the muon~\cite{DMR} or 
the earlier
LEP1, LEP-1.5 and Tevatron data. As more data is accumulated at Run-II of the 
Tevatron, we expect this bound to improve, but this improvement may be only 
marginal, as
Tevatron constraints are severely limited by the kinematic reach of the
machine. The commissioning of
high-energy machines like the CERN LHC and a possible linear $e^+e^-$ 
collider would 
increase the discovery reach for torsion fields enormously. A really high 
energy machine, like, for example, a muon collider running at 3--4~TeV 
center-of-mass energy,
would be enormously effective in such searches. Till such
data are available, however, we expect our results to more-or-less
represent the 
state-of-the-art for laboratory constraints on dynamical torsion scenarios.

\bigskip
\centerline{\bf Acknowledgments}

The authors acknowledge useful discussions with S.~Banerjee. SR would like
to further thank N.K.~Mondal and S.K.~Rai for discussions.
UM thanks the Department of Physics, Indian Institute of Technology, Kanpur 
for hospitality; SR likewise extends thanks to the Harish-Chandra Research 
Institute, Allahabad.

%%%%%%%%%%%%%%%%%%%%%%%%%%%%%%%%%%%%%%%%%%%%%%%%%%%%%%%%%%%%%%%%%%%%%%%%%%%

%%%%%%%%%%%%%%%%%%%%%%%%%%%%%%%%%%%%%%%%%%%%%%%%%%%%%%%%%%%%%%%%%%%%%%%%%%%

\begin{thebibliography}{99}
\bibitem{choice} We make no attempt, in this work, to motivate such a 
choice, which is clearly not unique.
We could, for example, have chosen opposite signs for all the 
couplings. We have checked that making that choice would lead to somehat
less restrictive constraints from Tevatron data. LEP-2 constraints would, 
of course, remain unaffected.
\bibitem{Bayesian} Such as, for example, the use of Bayesian statistics,
which are relevant when a few events are seen and are now employed by the
D0 Collaboration as a matter of course.
\end{thebibliography}
\end{document}